\documentclass[prl,twocolumn,showpacs,amsmath,amssymb]{revtex4}
\usepackage{graphicx}
\usepackage{dcolumn}
\usepackage{bm}
\begin{document}

\title{Solving the two-center nuclear shell-model problem with arbitrarily-orientated deformed potentials}

\author{Alexis Diaz-Torres}

\affiliation{Department of Nuclear Physics, Research School of
Physical Sciences and Engineering, Australian National University,
Canberra, ACT 0200, Australia}

\date{\today}

\begin{abstract}
A general new technique to solve the two-center problem with arbitrarily-orientated deformed realistic potentials is demonstrated, which is based on the powerful potential separable expansion method. As an example, molecular single-particle spectra for 
$^{12}$C + $^{12}$C $\to$ $^{24}$Mg are calculated using deformed Woods-Saxon potentials. These clearly show that non-axial symmetric configurations play a crucial role in 
molecular resonances observed in reaction processes for this system at low energy. 
\end{abstract}


\maketitle

The description of a particle moving in the field of two fixed potential centers separated by a distance $R$ (the two-center problem) is fundamental in 
classical \cite{Howard} and quantum mechanics \cite{Slater,Mueller}, and finds myriad applications in celestial mechanics, 
quantum chemistry, atomic and molecular physics and nuclear physics. This appears (i) in the study of the scattering of radiation by two black holes 
\cite{Chandrasekbar}, (ii) in the quantum mechanical theory of chemical binding 
\cite{Teller}, (iii) in the description of electron-positron pair production in heavy ion and ion-atom collisions \cite{Eichler}, (iv) in the study of the properties of baryons containing two heavy quarks (QQq) \cite{He}, and (v) in phenomena related to nuclear molecules \cite{GreinerParkScheid}. 

The applications to phenomena in low energy nuclear physics \cite{GreinerParkScheid} were first introduced (in practice) by the Frankfurt school using the two-center shell model (TCSM) based on a double oscillator potential \cite{Maruhn}. Improved versions of this approach have been suggested for dealing with super-asymmetric fission \cite{Mirea} and asymmetric fission with deformed fragments \cite{Radu}. In all these models, the two-center potentials are rotationally symmetric about the internuclear axis. These potentials are appropriate, e.g., for the description of binary fission where the fragments are spherical or deformed with their intrinsic symmetry axis aligned with the internuclear axis. 

All alignments are possible in collisions of deformed nuclei. The major role of orientation of the deformed target in the onset of quasi-fission, whose understanding is very important to unpuzzle the formation mechanism of superheavy elements \cite{Oganessian}, has been demonstrated \cite{David} by fission measurements in reactions forming heavy elements. A more general TCSM is required for a proper description of these reactions within the molecular picture \cite{GreinerParkScheid}. It is justified at low incident energies near the Coulomb barrier, as the radial motion of the nuclei is expected to be adiabatically slow compared to the rearrangement of the two-center mean field of nucleons. To my knowledge, only one attempt has been made to account for arbitrarily-orientated deformed nuclei, in which the wave function expansion method (usual diagonalization procedure) and specific potentials (two ellipsoidally deformed Gausssian potentials) were applied to describe the reaction $^{13}$C + $^{16}$O \cite{Nuhn}.  

I present a general new technique to solve the two-center problem with arbitrarily-orientated deformed fragments. The procedure is based on the powerful potential separable expansion method \cite{PSE} that has been successfully used to solve the two-center problem with spherical Woods-Saxon (WS) potentials \cite{Gareev,SPH_TCWS}. The technique is shown using deformed WS potentials, but it can also be employed with other types of deformed potential, provided a suitable set of basis functions is selected. Hence, the method has applications in many areas of science. The formalism is described first, and illustrated afterwards with calculations of molecular single-particle (sp) spectra for the reaction $^{12}$C + $^{12}$C, which is of great astrophysical interest \cite{Rolfs}. These calculations show the major role of non-axial symmetric configurations in forming a nuclear molecule, as the overlapping nuclei keep their identity and can ``dance" for quite some time at contact.  

\begin{figure}
\begin{center}
\includegraphics[width=8.0cm]{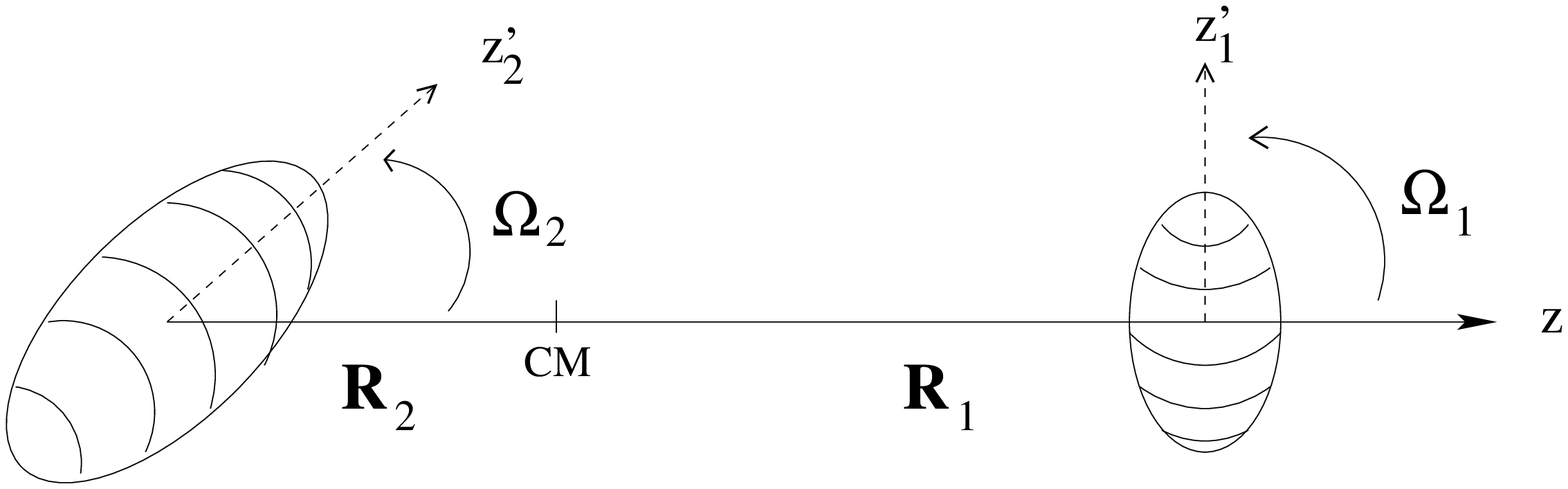}
\end{center}
\caption{Schematic picture of the coordinates used to define the two-center potential 
(\ref{eq1}) in the collision between two deformed nuclei. See text for further details.}
\label{Fig1}
\end{figure}

The finite depth nuclear potential $V_s(\textbf{r},\beta_{\lambda 0})$ of each spheroidal nucleus ($s=1,2$) is chosen to be a deformed WS 
\cite{Faessler} with a spin-orbit term and deformation parameters $\beta_{\lambda 0}$. For protons, the Coulomb potential $V_{Coul}^s(\textbf{r},\beta_{\lambda 0})$ \cite{Tamura} of a uniformly charged spheroid with charge $Z_s e$ ($Z_s$ being the total charge of each fragment) should be added to the nuclear potential. This treatment of the Coulomb interaction is suitable for separated fragments, but may not be the most appropriate prescription for overlapping nuclei. Its validity could be assessed by comparing this two-center Coulomb field to the field generated by a uniformly distributed charge within a dinuclear shape. The Coulomb interaction shifts up the proton levels with respect to the neutron levels, and increases the sp potential barrier between the fragments \cite{SPH_TCWS}. 

The origin of the total deformed potential $V_s(\textbf{r},\beta_{\lambda 0})$ is placed at the position $\textbf{R}_s$ in the overall center-of-mass (CM) system, where its intrinsic symmetry axis is orientated by Eulerian angles $\Omega_s =(\phi_s,\theta_s,0)$ ($0\leq \phi_s \leq 2\pi$ and $0\leq \theta_s \leq \pi$) with respect to the initial internuclear axis (see Fig.\ref{Fig1}). Thus, the two-center potential is
\begin{equation}
V=\sum_{s=1}^{2} e^{-i \textbf{R}_s \hat{k}}\, \hat{U}(\Omega_s)\ V_s\ 
\hat{U}^{-1}(\Omega_s)\, e^{i \textbf{R}_s \hat{k}}, 
\label{eq1}
\end{equation} 
where $\hat{k}=\hbar^{-1}\hat{p}$ is the sp wave-number operator and 
$\hat{U}$ is the operator of finite rotations \cite{Merzbacher}, whose inclusion is the key new aspect of this work. Of course, the rotation operator is not required for spherical nuclei \cite{SPH_TCWS}. Each potential $V_s$ in (\ref{eq1}) is expressed as 
\begin{equation}
V_{s} \approx \sum_{\nu \mu}^{N} |s\nu\rangle\, V_{\nu \mu}^s\, \langle s\mu|,
\label{eq1b}
\end{equation}
within a truncated spherical sp harmonic oscillator basis, 
$\{|\nu \rangle, \nu = 1, \ldots N \}$, with the spin-angular part 
having the total angular momentum $j$ with projection $m$, e.g., in the momentum 
representation (see \cite{SPH_TCWS} for further details)
\begin{equation}
|\nu\rangle = |nljm\rangle = g_{nl}(k)\cdot [i^{-l}Y_l (\hat{\bf{k}})
\otimes \chi_{\frac{1}{2}}(s) ]^{j}_{m}. 
\label{eq2}
\end{equation}

The number $N$ of basis states is defined by $l_{max}$ (number of partial waves in which 
the potential acts) and $n_{max}$ (the number of separable terms in each partial wave). The 
values of $l_{max}$ and $n_{max}$ are determined by the convergence of the sp energies, which is accelerated using the technique of the Lanczos $\sigma$-factors \cite{LanczosDiaz}. These are $n_{max}$=3 and $l_{max}$=4 for the studied reaction 
$^{12}$C + $^{12}$C. 

For bound (or quasi-stationary) states, the formal solution of the Schr\"odinger equation is
\begin{equation}
|\varphi\rangle = G_0(E) V |\varphi\rangle,
\quad G_0(E)= \biggl( E - \frac{\hbar^2\hat{k}^2}{2m_0}\biggl) ^{-1}, 
\label{eq3}
\end{equation}
where $G_0$ is the Green operator of the free sp motion. Inserting (\ref{eq1}) with 
(\ref{eq1b}) into (\ref{eq3}), and multiplying from the left by 
$\langle s\mu| \hat{U}^{-1}(\Omega_s) e^{i \textbf{R}_s \hat{k}}$, the following set of linear equations for the amplitudes 
$A_{s\mu}=\langle s\mu| \hat{U}^{-1}(\Omega_s) e^{i \textbf{R}_s \hat{k}}|\varphi\rangle$ is obtained:
\begin{eqnarray}
\sum_{\mu'=1}^{N}\sum_{s'=1}^{2} \biggl[ \delta_{ss'} \delta_{\mu \mu'} - 
\sum_{\nu=1}^{N} \langle s\mu| \hat{U}^{-1}(\Omega_s)\, G_0(E)e^{i \textbf{R}_{ss'}\hat{k}} \, \nonumber \\ \hat{U}(\Omega_{s'})|s'\nu \rangle
\, V_{\nu \mu'}^{s'} \biggl] A_{s'\mu'} = 0,  
\label{eq4}
\end{eqnarray}
where $\textbf{R}_{ss'}=\textbf{R}_s - \textbf{R}_{s'}$. Here there is no direct overlap between the two (non-orthogonal) set of basis functions, unlike in the secular matrix equation that results from the wave function expansion method \cite{Nuhn}. Only the off-diagonal block ($s \neq s'$) of the matrix for the linear system (\ref{eq4}) contains the dependence on the nuclei orientation $\Omega_s$, as 
$\hat{U}^{-1}(\Omega_s)\, \hat{U}(\Omega_s)$ is the unitary operator. 
The matrix elements in (\ref{eq4}) involving the Green operator $G_0$ are expressed in terms of the Wigner $D$-functions as follows 
\begin{eqnarray}
\sum_{m_1m_2} D^{j^*}_{m_1m}(\Omega_s)\, 
\langle snljm_1|G_0(E)e^{i \textbf{R}_{ss'}\hat{k}}|s'n'l'j'm_2\rangle\, 
 \nonumber \\ 
D^{j'}_{m_2m'}(\Omega_{s'}), \label{eq5}
\end{eqnarray}
where the new matrix elements in (\ref{eq5}) are explicitly given in Ref. \cite{SPH_TCWS} [see expressions (12)-(16)]. The off-diagonal term ($s \neq s'$) of these new matrix elements vanishes for large separations \cite{Gareev}, turning the two-center problem into two independent one-center problems associated with the individual deformed nuclei.

The system of algebraic equations (\ref{eq4}) is equivalent to the 
Lippmann-Schwinger equation (\ref{eq3}) with separable potentials (\ref{eq1b}), whose solution is exact. The solvability condition is that its determinant vanishes, leading to the adiabatic energies $E$ that are a parameter in the Green operator $G_0$. With the eigenvalues $E$, the eigenstates $|\varphi\rangle$ are obtained solving the system (\ref{eq4}) for the amplitudes $A_{s \mu}$ and requiring the normalization of the state vectors $|\varphi\rangle$. For well-separated nuclei, the eigenstates are those of the individual nuclei, being the projection of the sp total angular momentum along their intrinsic symmetry axis a good quantum number. Molecular orbitals develop at small separations, which may not have good quantum numbers (there are no symmetries) as the two set of basis states are completely mixed by the potential and the operator of finite rotations. For identical potentials (mutually aligned identical nuclei), a symmetry of the two-center potential (\ref{eq1}) arises, namely its invariance under the permutation of the individual orientated potentials with respect to the CM point in Fig.\ref{Fig1}. In this case, the parity of the molecular sp states is also a good quantum number. Symmetric and anti-symmetric linear combinations of those (asymptotically degenerated) states, with opposite parity, result in (atomic) states localized around one of the potentials \cite{Gareev}. Where this is not the case, the solution of the two-center problem at large separations directly yields states of the individual nuclei.

\begin{figure}
\begin{tabular}{cc}
\includegraphics[width=0.20\textwidth,angle=0]{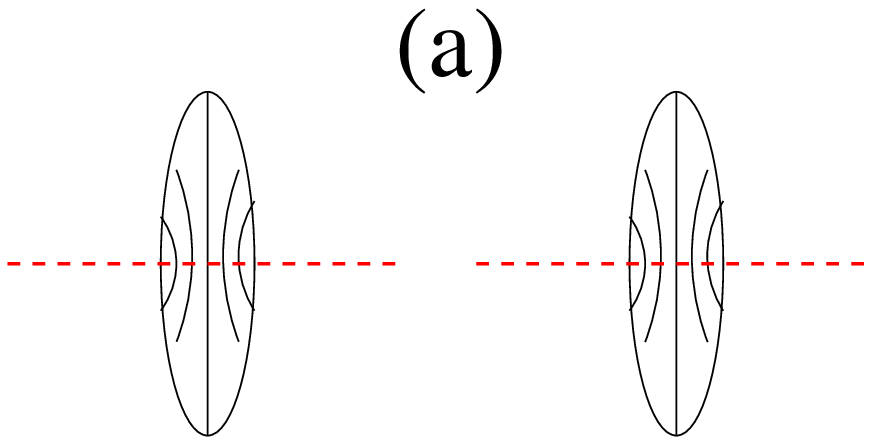} &
\includegraphics[width=0.20\textwidth,angle=0]{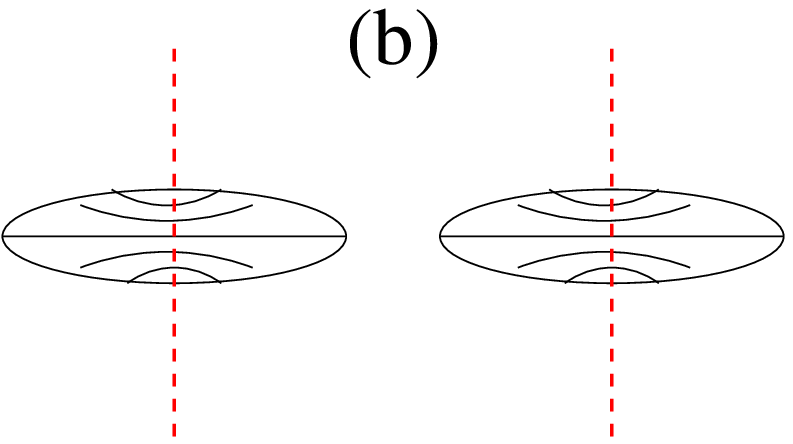} \\
\includegraphics[width=0.20\textwidth,angle=0]{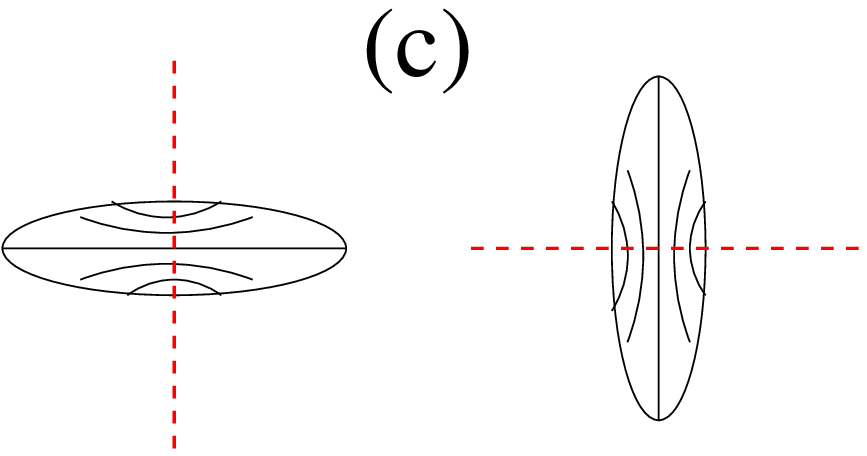} &
\includegraphics[width=0.20\textwidth,angle=0]{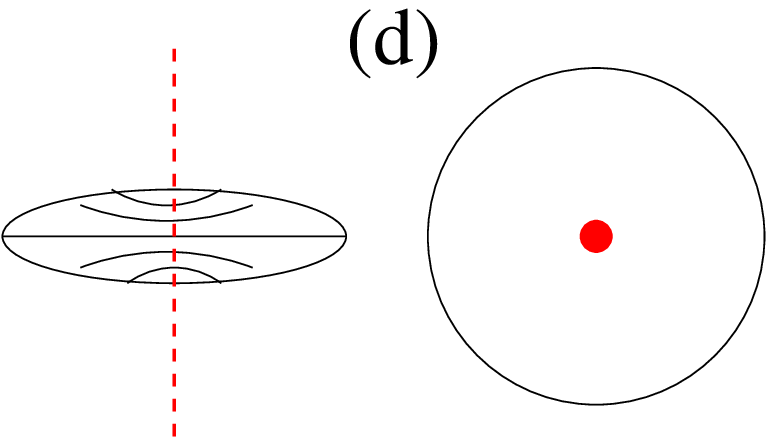}
\end{tabular}
\caption{Different configurations for the reaction $^{12}$C + $^{12}$C: axial symmetric (a), and non-axial symmetric (b)-(d).}
\label{orientations}
\end{figure}  

\begin{figure}
\begin{tabular}{cc}
\includegraphics[width=0.24\textwidth,angle=0]{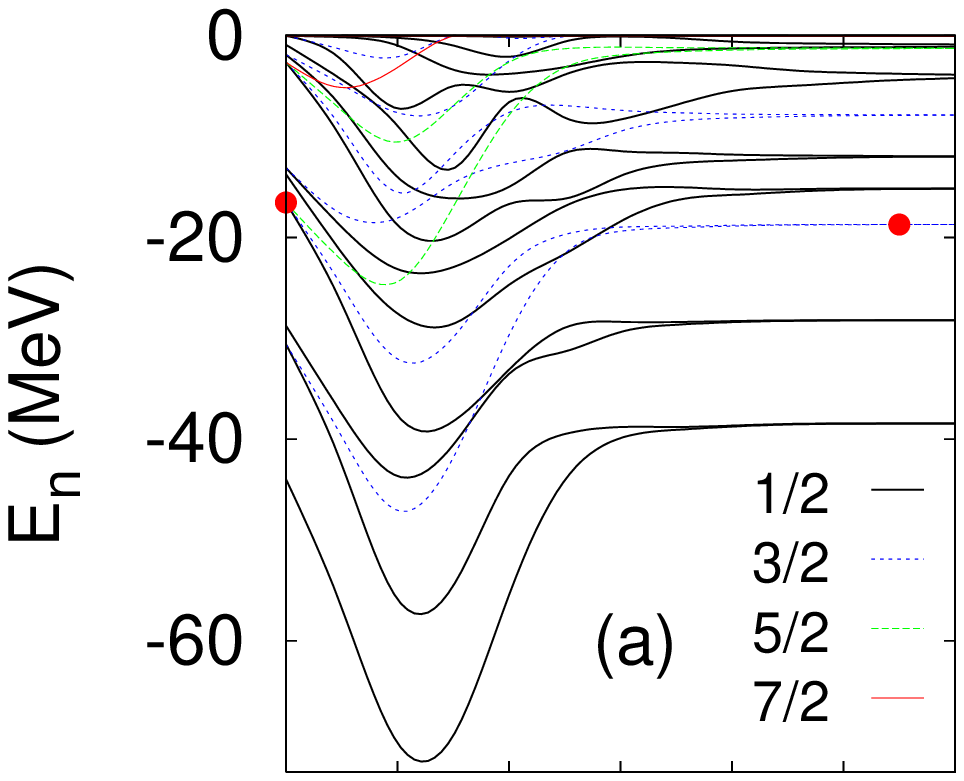} &
\includegraphics[width=0.24\textwidth,angle=0]{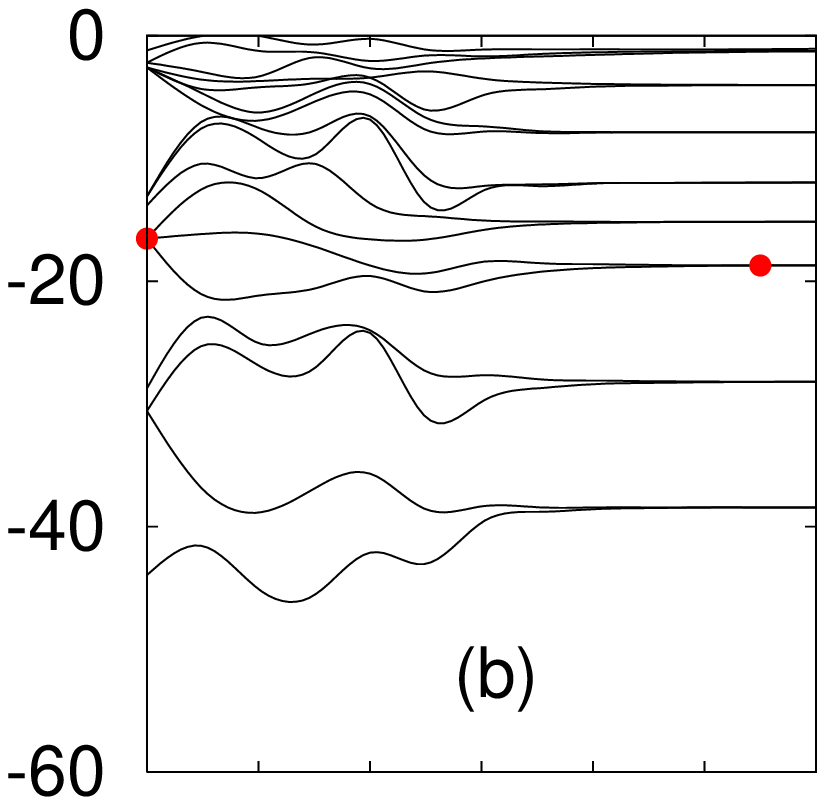} \\
\includegraphics[width=0.24\textwidth,angle=0]{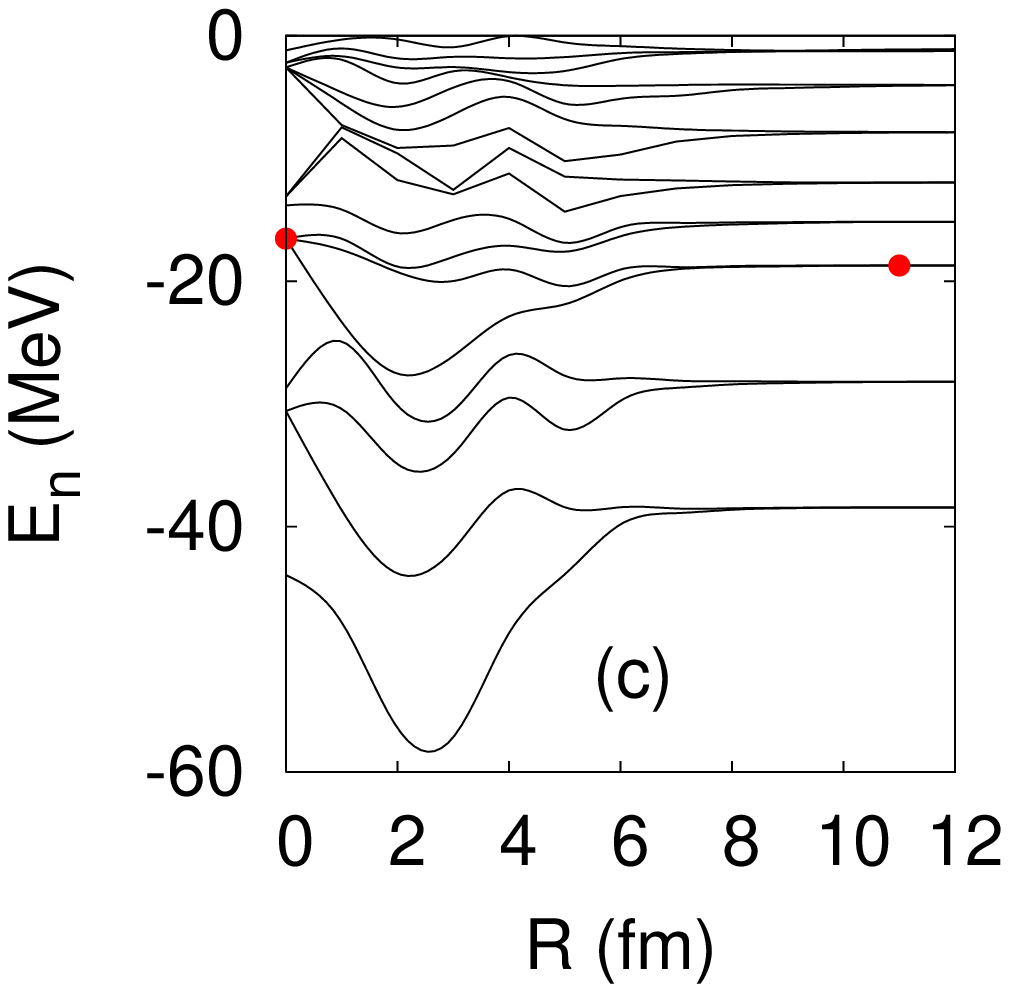} &
\includegraphics[width=0.24\textwidth,angle=0]{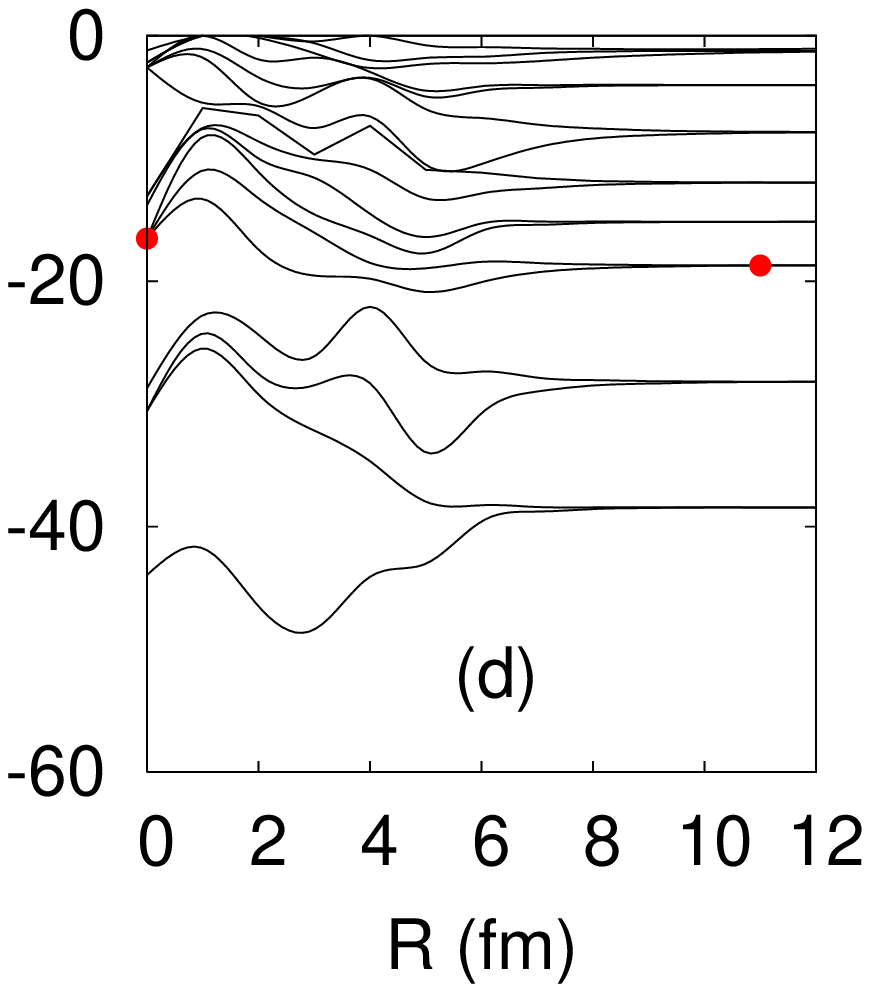}
\end{tabular}
\caption{(Color online) Neutron molecular sp levels as a function of the internuclear distance $R$, corresponding to the configurations shown in Fig.\ref{orientations}. 
Different curves in panel (a) are associated with different magnetic quantum numbers. 
The points denote the Fermi level of the spherical compound nucleus and the $^{12}$C 
nuclei.}
\label{neutron_spectra}
\end{figure}

\begin{figure}
\begin{tabular}{cc}
\includegraphics[width=0.24\textwidth,angle=0]{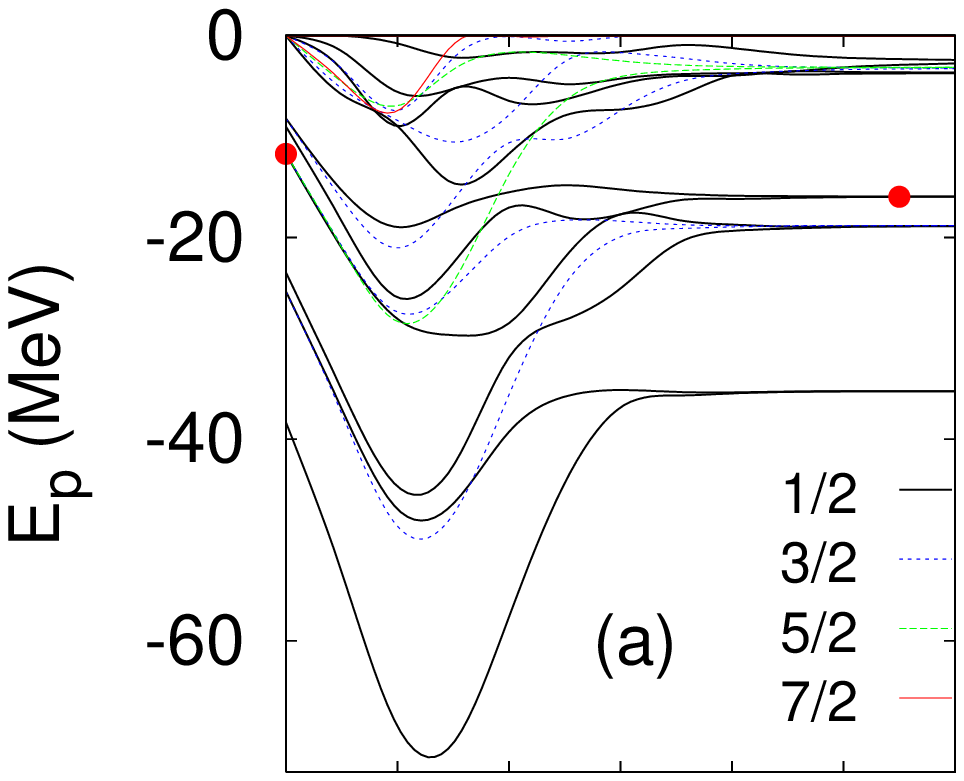} &
\includegraphics[width=0.24\textwidth,angle=0]{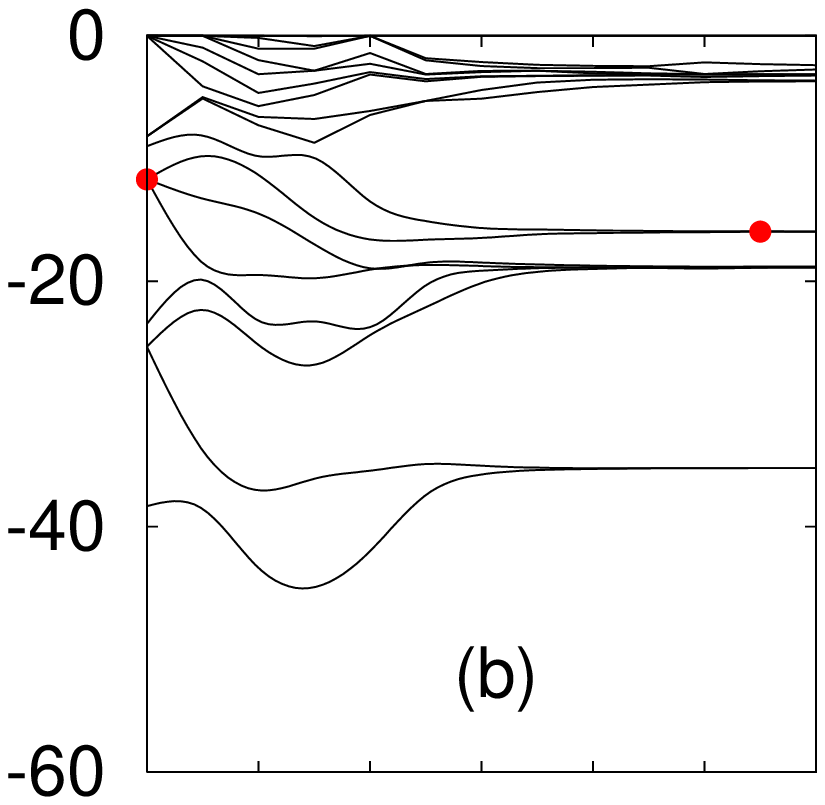} \\
\includegraphics[width=0.24\textwidth,angle=0]{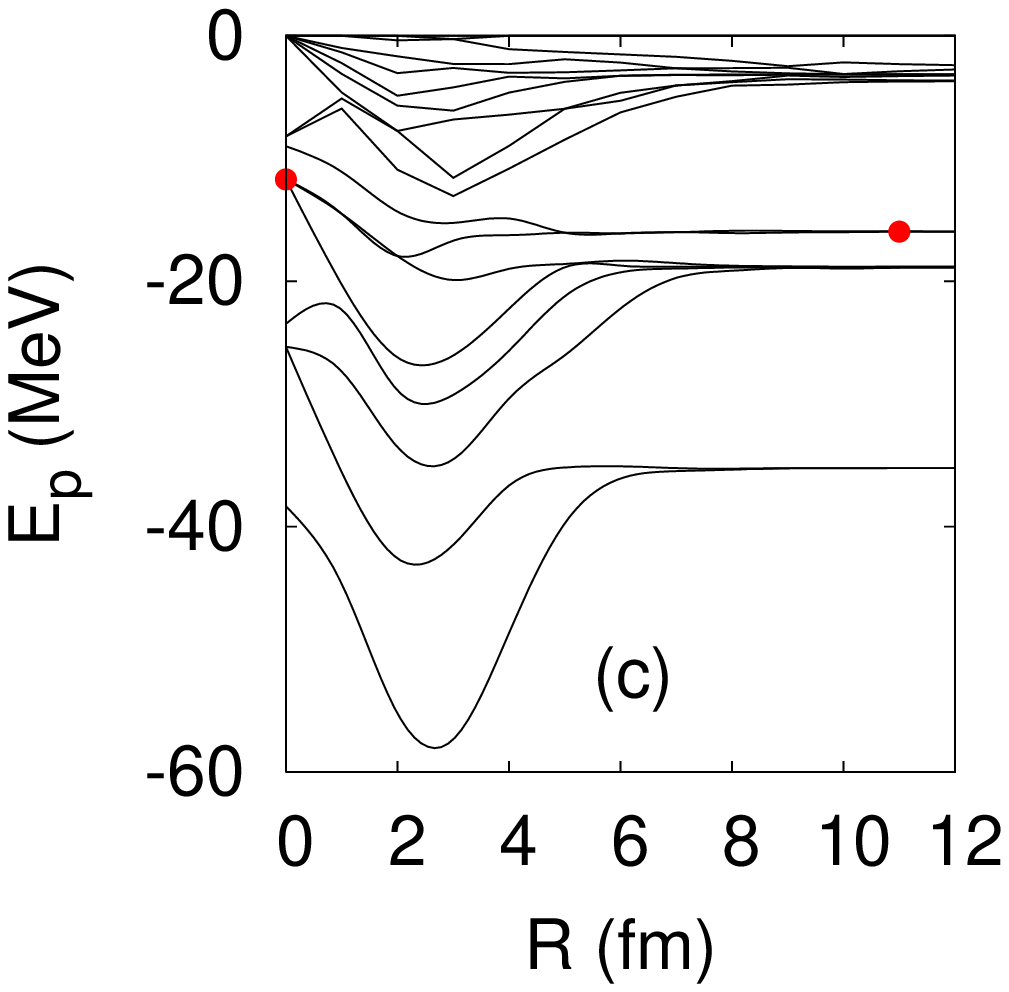} &
\includegraphics[width=0.24\textwidth,angle=0]{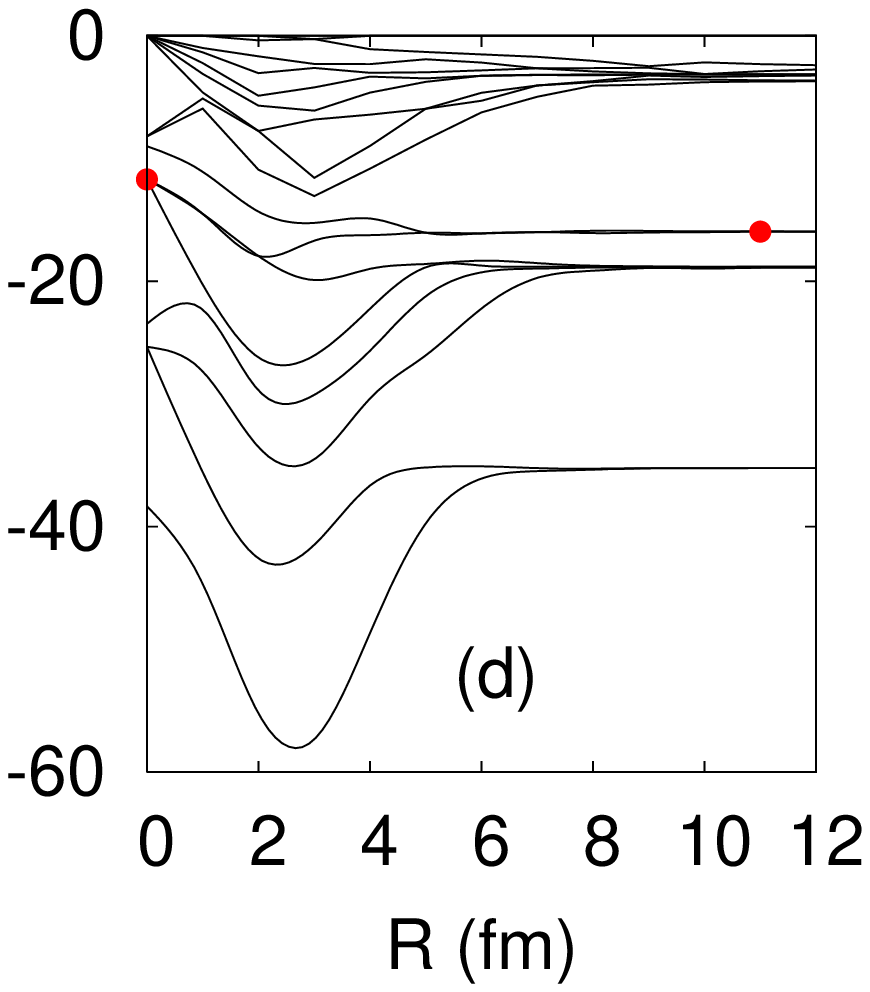}
\end{tabular}
\caption{(Color online) The same as Fig.\ref{neutron_spectra}, but for protons.}
\label{proton_spectra}
\end{figure}   

The formalism is now illustrated with calculations of molecular sp levels diagram (i.e., energy levels as a function of the internuclear distance $R$) for $^{12}$C + $^{12}$C $\to$ $^{24}$Mg, where $^{12}$C is an oblately deformed nucleus with quadrupole deformation $\beta_2 = -\, 0.5$ \cite{Lebedev}. Calculations are carried out for the collisions shown in Fig.\ref{orientations}. The parameters of the asymptotic WS potentials ($R=0$ and $R \to \infty$) including the spin-orbit term are the global parameters by Soloviev \cite{Soloviev}, where the depth of the potentials has been adjusted to reproduce the experimental sp separation energies \cite{Audi}. To describe fusion, all the potential parameters (including those of the Coulomb potential for protons) have to be interpolated between their values for the separated nuclei and the spherical compound nucleus \cite{SPH_TCWS}. The parameters can be correlated by conserving the volume enclosed by a certain equipotential surface (the Fermi level of the spherical fused system) of the two-center potential (\ref{eq1}), for all separations $R$ between the nuclei with orientations $\Omega_s$ \cite{Nuhn,SPH_TCWS}. The nuclear shape is considered to be the same for neutrons and protons, with the neck size naturally determined by the superposition of the smooth tail of the two WS potentials. However, a variable neck size could be included in the method through an additional intermediate potential, resulting in a three-center problem. This may be very useful in fission studies and cluster physics.

Figures \ref{neutron_spectra} and \ref{proton_spectra} show the molecular sp levels diagram for neutrons and protons, respectively, for different orientations of the two oblately deformed $^{12}$C, as presented in Fig.\ref{orientations}. Only for the aligned orientation of the deformation axis with the internuclear axis [axial symmetric configuration in panel (a)], the projection of the nucleon total angular momentum along the internuclear axis is a good quantum number at all separations, whose values are represented by different curves. Here as in panel (b), the parity of the molecular orbitals (not indicated) is also a good quantum number, whose values are easily deduced from the orbital angular momentum $l$ of the sp levels of the spherical compound nucleus ($R=0$), i.e., $(-1)^l$. The full circles denote the Fermi level of the $^{12}$C nuclei and the spherical compound nucleus. As expected, the asymptotic shell structure does not depend on the mutual alignment of the $^{12}$C nuclei.  

These molecular spectra show significant features:
\begin{itemize}
\item [(i)] The asymptotic shell structure of $^{12}$C is much less distorted for non-axial symmetric configurations [panels (b)-(d)] than for the axial symmetric one [panel (a)], the former showing more preservation of the indentity of the overlapping nuclei.
\item [(ii)] For non-axial symmetric configurations, the neutron levels show a minimum at separations between 4-6 fm, which may result in a ``molecular pocket" in the collective potential energy surface \cite{GreinerParkScheid}. This depends weakly on the mutual alignment.
\item [(iii)] Many avoided crossings appear between 2-6 fm, in which the sp wave-function abruptely changes its nodal structure. It may lead to strong peaks in the radial collective mass parameter \cite{16O16O}, which can hinder the fusion of the nuclei. The critical radius for fusion \cite{Mosel1} (where the shell structure of $^{24}$Mg starts to build up) is quite small for non-axial symmetric configurations ($\lesssim$ 2 fm), which also favours the formation of a nuclear molecule (the weakly overlapping nuclei remain longer around 4-6 fm). 
\end{itemize}

The preservation of the identity of the nuclei along with their being trapped around the contact distance are crucial aspects for the formation of a nuclear molecule \cite{GreinerParkScheid}. These favourable features are shown by non-axial symmetric configurations of 
$^{12}$C + $^{12}$C. Since their shell structures are quite similar, the $^{12}$C orientation is clearly an essential variable in the reaction processes. This degree of freedom activates \cite{Hess1,Abe1} dynamical modes (butterfly, anti-butterfly, belly-dancer etc) at contact, making the nuclei ``dance" there for quite some time. It should result in narrow resonances in the reaction cross sections, as shown by measurements \cite{Aguilera}. The molecular sp spectra are useful to microscopically obtain collective potentials and mass surfaces for a molecular reaction dynamical calculation \cite{Abe1}. It is worth mentioning that other studies \cite{Chandra,Tanimura,Niklas,Schmidt} have also argued the importance of the non-axial symmetric configurations for effects of molecular resonances on reaction processes for $^{12}$C + $^{12}$C. 

In summary, a general new technique to solve the two-center problem with arbitrarily-orientated deformed realistic potentials has been demonstrated. Among other applications such as in cluster physics \cite{Freer}, this should be very useful for describing, within the molecular picture, low energy nuclear reaction processes involving deformed nuclei, such as (i) formation of heavy and superheavy elements \cite{Alexis2}, (ii) effects of breakup of weakly-bound nuclei on fusion \cite{Alexis3}, and fusion reactions of great astrophysical interest \cite{16O16O}. Molecular sp spectra clearly show that non-axial symmetric configurations are crucial for the formation of a nuclear molecule in the reaction $^{12}$C + $^{12}$C. Reaction dynamical calculations for a quantitative understanding of molecular resonance structures in its astrophysical S-factor \cite{Aguilera} are in progress.

%\begin{center}
%{\bf AKNOWLEDGEMENTS}
%\end{center}

The author thanks Prof. W. Scheid for discussions. Support from an ARC Discovery grant is acknowledged.

\end{document}